\documentclass[aps,preprint,prd,showpacs,nofootinbib]{revtex4}
\usepackage{latexsym}
\usepackage{CJK}
\usepackage{amsmath}
\usepackage{graphicx}
\usepackage{subfigure}
\usepackage{dcolumn}
\usepackage{bm}
\usepackage{amssymb}
\usepackage{latexsym}
\usepackage{ulem}
\usepackage{color}
\usepackage[colorlinks,linkcolor=magenta,anchorcolor=blue,citecolor=blue]{hyperref}

\usepackage{extpfeil}
\usepackage{extarrows}
\usepackage{boxedminipage}

\def\be{\begin{equation}}
\def\ee{\end{equation}}
\def\ba{\begin{eqnarray}}
\def\ea{\end{eqnarray}}

\def\nn{\nonumber}
\def\lf{\left}
\def\rt{\right}

\input{epsf.sty}

\begin{document}

\title{The Effective Field Theory of nonsingular cosmology}

\author{Yong Cai$^{1}$\footnote{caiyong13@mails.ucas.ac.cn}}
\author{Youping Wan$^{1,2}$\footnote{wanyp@ustc.edu.cn}}
\author{Hai-Guang Li$^{1}$\footnote{lihaiguang14@mails.ucas.ac.cn}}
\author{Taotao Qiu$^{3,5}$\footnote{qiutt@mail.ccnu.edu.cn}}
\author{Yun-Song Piao$^{1,4}$\footnote{yspiao@ucas.ac.cn}}

\affiliation{$^1$ School of Physics, University of Chinese Academy
of Sciences, Beijing 100049, China}
\affiliation{$^2$ CAS Key Laboratory for Research in Galaxies and Cosmology,
Department of Astronomy, University of Science and Technology of China,
Chinese Academy of Sciences, Hefei, Anhui 230026, China}
\affiliation{$^3$ Institute of Astrophysics, Central China Normal University, Wuhan 430079, China}
\affiliation{$^4$ Institute of Theoretical Physics, Chinese
Academy of Sciences, P.O. Box 2735, Beijing 100190, China}
\affiliation{$^5$ Key Laboratory of Quark and Lepton Physics (MOE), Central China Normal University, Wuhan 430079, P.R.China}


\begin{abstract}

In this paper, we explore the nonsingular cosmology within the
framework of the Effective Field Theory(EFT) of cosmological
perturbations. Due to the recently proved no-go theorem, any
nonsingular cosmological models based on the cubic Galileon
suffer from pathologies. We show how the EFT could help us
clarify the origin of the no-go theorem, and offer us solutions
to break the no-go. Particularly, we point out that the gradient
instability can be removed by using some spatial derivative
operators in EFT. Based on the EFT description, we obtain a
realistic healthy nonsingular cosmological model, and show
the perturbation spectrum can be consistent with the
observations.

\end{abstract}

\maketitle

\section{Introduction}

In the modern era of Cosmology, theories of Hot Big Bang (HBB) and
Inflation have achieved great success, and thus have long been regarded as
the standard paradigm of the early universe. However, the
inflation still suffers from the cosmological singularity problem
\cite{Borde:1993xh}\cite{Borde:2001nh}, unless it was preceded by
a bounce \cite{Piao:2003zm}\cite{Liu:2013kea}\cite{Qiu:2015nha} or a Genesis phase
\cite{Liu:2014tda}\cite{Pirtskhalava:2014esa}\cite{Kobayashi:2015gga}.
It is exciting to study classical nonsingular cosmology, such as bounce universe
models \cite{Cai:2007qw}\cite{Cai:2008qw}, Genesis models
\cite{Creminelli:2010ba}\cite{Hinterbichler:2012yn}\cite{Cai:2013rna},
slow expansion models \cite{Piao:2003ty}\cite{Liu:2011ns}\cite{Liu:2012ww},
since we might get classical nonsingular cosmology without
begging the details of the unknown UV-complete gravity theory.

One of the most exciting endeavors in this area is to build nonsingular
cosmological models with the field theories which can violate the Null Energy
Condition (NEC) \cite{Rubakov:2014jja}. Usually the violation of NEC
may lead to the ghost instability \cite{Carroll:2003st}. This problem can be solved
if one considers the so-called Galileon theory \cite{Nicolis:2008in} or its
generalizations (such as the Horndeski theory \cite{Horndeski:1974wa}\cite{Deffayet:2011gz}
and its beyond \cite{Gleyzes:2014dya}).  Making
use of the simplest cubic Galileon, many heuristic nonsingular
cosmological models have been constructed, e.g. \cite{Creminelli:2010ba}\cite{Liu:2011ns}\cite{Qiu:2011cy}\cite{Qiu:2013eoa}.
However, it seems quite difficult to
avoid the gradient instability problem, which indicates a negative sound
speed squared \cite{Qiu:2015nha}\cite{Pirtskhalava:2014esa}\cite{Cai:2012va}\cite{Wan:2015hya}
and thus leads to an exponential growth of the perturbation
\cite{Battarra:2014tga}\cite{Koehn:2015vvy}.

Recently, Libanov, Mironov and Rubakov (LMR) have proved a no-go theorem,
which shows that healthy nonsingular cosmological models based on the cubic
Galileon does not exist \cite{Libanov:2016kfc}. Hereafter, it was generalized
with an additional scalar in Ref.\cite{Kolevatov:2016ppi} or with
the full Horndeski theory in Ref.\cite{Kobayashi:2016xpl}. However,
Ijjas and Steinhardt claimed that there exists a loophole in the proof of
Ref.\cite{Kobayashi:2016xpl} (which was also noticed by the author of
Ref.\cite{Kobayashi:2016xpl}), and they can even reconstruct a fully stable
classical bounce \cite{Ijjas:2016vtq} throughout the whole evolution by using
the ``inverse method" \cite{Ijjas:2016tpn}. However, we believe that this relevant
issue still needs to be studied further.

Prior to LMR's work, studies were also made along other lines. The danger
of $c_s^2<0$ is mainly attributed to the exponential growth of the amplitude of
short wavelength modes. In \cite{Battarra:2014tga}\cite{Koehn:2015vvy}, it was
argued that the strong coupling scale during $c_s^2<0$ is low so that the
dangerous short wavelength modes lie outside the range of the validity of the
effective theory, thus can be disregarded. However, this argument begs unknown
strong coupling physics, which actually makes the ``classical nonsingular" bounce
loose sense. What is the effective theory of nonsingular cosmology is a significant
issue. It is interesting to notice that some spatial covariant operators also help to
remove the gradient instability \cite{Qiu:2015nha}\cite{Pirtskhalava:2014esa}\cite{Kobayashi:2015gga}.

The Effective Field Theory (EFT) of cosmological perturbations is
extremely powerful and has been widely used to  study inflation
\cite{Cheung:2007st}\cite{Weinberg:2008hq} and dark energy
\cite{Gubitosi:2012hu}\cite{Gleyzes:2013ooa}\cite{Piazza:2013coa}.
It offers a unifying platform to deal with the cosmological perturbations
of all kinds of theories, such as the Horndeski theory and its beyond,
the Horava gravity \cite{Kase:2014cwa}, and the spatial covariant
gravity\cite{Gao:2014soa}\cite{Gao:2014fra}. In the following context,
we will see that it is also a powerful tool for studying nonsingular
cosmology.

In this paper, we will explore how to build healthy nonsingular
cosmological models within the framework of EFT. Practically, in
Sec.\ref{no-go}, based on EFT, we clarify how to understand
the no-go theorem and how to avoid it. We find that some effective
operators can play significant role in building nonsingular
cosmological models without pathologies. In
Sec.\ref{EFT-Perturbation}, we study the evolution of primordial
perturbation in nonsingular models with these corresponding
effective operators, and find the perturbation spectrum can be
consistent with the observations.  In Sec.\ref{Model}, we present
a realistic healthy nonsingular bounce model by introducing an
effective operator of $R^{(3)}\delta g^{00}$. Finally, we conclude
in Sec.\ref{Conclusion}.

\textbf{Note added:} After our paper appeared in arXiv, nearly simultaneously Creminelli sent us their draft (the preprint \cite{Creminelli:2016zwa}), which overlaps substantially with ours.

\section{ the framework of EFT and the no-go theorem}
\label{no-go}

We consider the metric in the ADM form:
\be
\label{metric}
ds^2=-N^2dt^2+h_{ij}(dx^i+N^idt)(dx^j+N^jdt)~,
\ee
where $N$ and $N^i$ are the lapse function and shift vector, and $h_{ij}$ is
the 3-dimentional spatial metric.

With the spirit of the EFT of cosmological perturbation
\cite{Cheung:2007st}\cite{Gubitosi:2012hu}\cite{Kase:2014cwa}, we write down the EFT action
for nonsingular cosmological models
\ba
\label{eft_action}
S&=&\int d^4x\sqrt{-g}\Big[
{M_p^2\over2} f(t)R-\Lambda(t)-c(t)g^{00}
\nn\\
&\,&+{M_2^4(t)\over2}(\delta g^{00})^2-{m_3^3(t)\over2}\delta
K\delta g^{00} -m_4^2(t)\lf( \delta K^2-\delta K_{\mu\nu}\delta
K^{\mu\nu} \rt) +{\tilde{m}_4^2(t)\over 2}R^{(3)}\delta g^{00}
\nn\\
&\,&-\bar{m}_4^2(t)\delta K^2+{\bar{m}_5(t)\over 2}R^{(3)}\delta K
+{\bar{\lambda}(t)\over2}(R^{(3)})^2+...
\nn\\
&\,& -{\tilde{\lambda}(t)\over
M_p^2}\nabla_iR^{(3)}\nabla^iR^{(3)} +... \Big] \,,
\ea
where we turn off
the accelerator vectors $a_i$ in \cite{Kase:2014cwa} for simplicity.
We assume the
matter part is minimally coupled to field so that the expansion
or contraction of the background with respect to physical rulers is unambiguous.
The first line describes the background of our model, while
the rest is for perturbations. One is also allowed to contain terms
such as ${R^{(3)}}_{\mu\nu}{R^{(3)}}^{\mu\nu}$ and $\nabla_i
{R^{(3)}}_{jk}\nabla^i {R^{(3)}}^{jk}$,  which we don't bother
to write them explicitly and just put them into the ellipsis. All
the coefficients are allowed to vary with $t$, with the
dimension $[m_i]=1$, $[\lambda_i]=0$, so as to make the
action dimensionless. Moreover, in this action we define
$\delta K_{\mu\nu}=K_{\mu\nu}-H H_{\mu\nu},~\delta K=K-3H$ ,
with the induced metric $H_{\mu\nu}\equiv g_{\mu\nu}+n_\mu n_\nu$
and the normal vector is defined as $n_\mu\equiv(-N,0,0,0)$.

It is rather straightforward to fix the relations among the functions
$f(t)$, $c(t)$ and $\Lambda(t)$, which is in the background part.
Varying the first line of action Eq.(\ref{eft_action}) with respect
to $N$ and $a$, one can get the two equations:
\ba
3M_p^2[f(t)H^2+\dot f(t)H]&=&c(t)+\Lambda(t)~,\\
-M_p^2[2f(t)\dot H+3f(t)H^2+2\dot f(t)H+\ddot
f(t)]&=&c(t)-\Lambda(t)~.
\ea
For the minimal coupling theories where $f(t)=1$, these are nothing
but the Friedmann equations, thus we have $c(t)=-M_p^2{\dot H}$
and $\Lambda(t)=M_p^2({\dot H}+3H^2)$. The $c(t)$ and $\Lambda(t)$
have the same expressions as those in the EFT of inflation, however, to have
a non-singular scenario a crucial condition must be satisfied, i.e., the violation
of NEC. That means $c(t)$ must be negative at least for a while. Since the
NEC will finally be restored in the expanding universe, we conclude for the
EFT of nonsingular cosmology, {\it $c(t)$ must be a function that can pass
the zero boundary}. For the case with non-minimal coupling, $f(t)$ is
nontrivial, then a more complicated constraint will be imposed on $c(t)$
and $\Lambda(t)$.

\subsection{The no-go theorem}

It is straightforward to derive the quadratic action of scalar and tensor
perturbation from Eq.(\ref{eft_action}). We give some main steps of the
derivation in Appendix \ref{appA} and just write down the results here.
Under the unitary gauge, the quadratic action of scalar perturbation is
\ba
\label{eft_action02}
 S^{(2)}_\zeta=\int d^4xa^3\lf[
c_1\dot{\zeta}^2-\lf({\dot{c}_3\over a} -c_2\rt){(\partial
\zeta)^2\over a^2}+{c_4\over a^4}(\partial^2\zeta)^2
-{16\tilde{\lambda}(t)\over M_p^2 a^6}\lf(\partial^3\zeta \rt)^2
\rt]\,,
\ea
where we have left the expressions of $c_{i}$ in Appendix \ref{appA}
since they are complicated (except for the $c_2$, which has a quite simple
expression as $c_2=M^2_p f(t)$). The sound speed squared reads
\begin{eqnarray}
\label{cs2}
c^2_s=\lf({\dot{c}_3\over a}-c_2 \rt)/c_1~.
\end{eqnarray}
The conditions to avoid the ghost instability and the gradient
instability are
\begin{eqnarray}
c_1 >0~,~~~\dot{c}_3 -ac_2>0~.
\end{eqnarray}
Moreover, the quadratic action of tensor perturbation from Eq. (\ref{eft_action}) is
\ba
S^{(2)}_{\gamma}={M_p^2\over8}\int d^4xa^3 Q_T\lf[
\dot{\gamma}_{ij}^2 -c_T^2{(\partial_k\gamma_{ij})^2\over
a^2}\rt]\,, \label{tensor-action}
\ea
where
\be
Q_T= f+2\lf({m_4\over M_p} \rt)^2\,, \qquad c_T^2={f\over  Q_T  }\,.
\ee
To avoid the ghost and gradient instability for tensor modes, we need
$Q_T>0$ and $c_T^2>0$, respectively.

We begin with $\dot{c}_3 -ac_2>0$, which indicates
\begin{eqnarray}
 c_{3}\big{|}_{t_f}-c_3\big{|}_{t_i}>\int^{t_f}_{t_i}ac_2 dt=M_p^2\int^{t_f}_{t_i} a f(t)dt~.
\label{k2-correction}
\end{eqnarray}
This expression is the key inequality to clarify the no-go theorem.
This inequality turns out to be remarkablely general since it is correct not
only for the Horndeski theory, but also for these theories beyond the
Horndeski. As matter of fact, by mapping the cubic Galileon to the EFT
\cite{Gleyzes:2013ooa}, Eq.(\ref{k2-correction}) will lead to the key
inequality used to prove the LMR no-go theorem \cite{Libanov:2016kfc}
(see the following part of this subsection); and by mapping the whole
Horndeski theory to the EFT \cite{Gleyzes:2013ooa}, Eq.(\ref{k2-correction})
will produce the key inequality in Kobayashi's paper \cite{Kobayashi:2016xpl}.

Now let's consider the cubic Galileon ${\cal L}_2+{\cal L}_3$ with $f(t)=1$,
Eq.(\ref{k2-correction}) reads
\begin{eqnarray}
\label{k2-correction-1}
 c_{3}\big{|}_{t_f}-c_3\big{|}_{t_i}>\int^{t_f}_{t_i}ac_2 dt=M_p^2\int^{t_f}_{t_i} a dt~,
\end{eqnarray}
and according to the Appendix. \ref{appA}, we find
\be c_3=
{2aM_p^4\over 2HM_p^2-m^3_3}={aM_p^2\over \gamma }~,
\ee
where $\gamma= H-(1/2)m^3_3/M_p^2$. We see from Eq.(\ref{k2-correction})
that $c_3$ is increased with time. Supposing $c_3\big{|}_{t_i}<0$, from
\begin{eqnarray}
c_3\big{|}_{t_f}>c_3\big{|}_{t_i}+M_p^2\int^{t_f}_{t_i}a dt
\label{sigma1}
\end{eqnarray}
we can tell that $c_3\big{|}_{t_f}$ will finally be larger than zero, thus
$c_3$ must equal to zero at sometime $t$ with $t_i<t<t_f$, making
$\gamma$ blows away. Therefore the gradient instability cannot be avoided.
The remaining case is that $c_3$ be always positive. However, from
\begin{eqnarray}
c_3\big{|}_{t_f}-M_p^2\int^{t_f}_{t_i}a  dt>c_3\big{|}_{t_i}
\end{eqnarray}
and let $t_i\rightarrow-\infty$, we see this is impossible in a
similar manner. So we have reformulated the LMR no-go theorem
\cite{Libanov:2016kfc} for the cubic Galileon in the framework of
EFT, which indicates the pathologies in nonsingular cosmological
models based on the cubic Galileon are inevitable.

It is interesting to note that Eq.(\ref{tensor-action}) can be
reformulated as \ba S^{(2)}_{\gamma}={{M_p^2}\over8}\int
dt_Ed^3xa^{3}_E\lf[\lf({\partial{\gamma}_{ij}\over \partial
t_E}\rt)^2 - {(\partial\gamma_{ij})^2\over a^{2}_E}\rt]\,
\label{tensor-action-E} \ea after a disformal redefinition of the
metric. Here we have defined $a_E=c_2^{1/2}(c_T^{-1/2}a)$ and
$dt_E=c_2^{1/2}(c_T^{1/2}dt)$, see e.g. Ref.\cite{Cai:2016ldn}.
This suggests \be \int^{t_f}_{t_i}ac_{2} dt =
\int^{t_{E,f}}_{t_{E,i}}a_E dt_E. \label{EF} \ee In certain sense,
the inequality Eq.(\ref{k2-correction}) is actually equivalent to
Eq.(\ref{k2-correction-1}). The integral $\int^{t_f}_{t_i}af dt$
(noting $c_2=M_p^2f$) corresponds to the affine parameter of the
graviton geodesics.

\subsection{How to evade the no-go theorem within the framework of EFT}
\label{evadenogo}


Recently, the no-go proof has been extended to the full Horndeski
theory by T. Kobayashi \cite{Kobayashi:2016xpl}. However, it seems that
this no-go theorem might be broken if the integral $\int^{t_f}_{t_i}af dt$
is not divergent \footnote{Note that in EFT description of Horndeski theory,
we have $f=2[G_4-X(\ddot\phi G_{5,X}+G_{5,\phi})]$ \cite{Gleyzes:2013ooa}
\cite{Kase:2014cwa}.}. Very recently, A. Ijjas and P. J. Steinhardt
found a fully stable bounce by keeping the integral $\int^{t_f}_{t_i}af dt$
convergent \cite{Ijjas:2016vtq}\cite{Ijjas:2016tpn}. In this section, we discuss how to avoid the no-go theorem within the
framework of EFT Eq.(\ref{eft_action}), while we assume $\int^{t_f}_{t_i}af dt$ is
divergent and $Q_T>0$ throughout (see \cite{Ijjas:2016vtq} for the cases $\int^{t_f}_{t_i}af dt$ is convergent or
$Q_T=0$ at some time),
which actually indicates that we have to go beyond Horndeski theory.


We firstly consider the addition of the effective operator
$R^{(3)}\delta g^{00}$ to the cubic Galileon.
It gives a contribution with $(\partial\zeta)^2\sim k^2\zeta_k^2$ to the
scalar perturbation, while does not change the tensor perturbation at
quadratic order.
The EFT action is written as:
\ba
S_{eff}
&=&\int d^4x\sqrt{-g}\Big[ {M_p^2\over2}R-\Lambda(t)-c(t)g^{00}
\nn\\
&\,&+{M_2^4(t)\over2}(\delta g^{00})^2-{m_3^3(t)\over2}\delta
K\delta g^{00} +{\tilde{m}_4^2(t)\over 2}R^{(3)}\delta
g^{00}\Big]~,
\label{k2-correction1}
\ea
here we have set $f(t)=1$, the coefficients $c(t)$, $\Lambda(t)$, $M_2^4(t)$ and $m_3^3(t)$ can be found by requiring that they have the same time-dependent behaviors as in the cubic Galileon ${\cal L}_2+{\cal L}_3$. The
existence of the last $\tilde{m}_4^2(t)$ term indicates this model
Eq.(\ref{k2-correction1}) goes beyond the Horndeski.
Note, the dynamical equation for the true degree of
freedom is still second order, thus the $\tilde{m}_4^2(t)$ term here,
as well as the higher order spacial derivative terms $(R^{(3)})^2$ and
$\nabla_iR^{(3)}\nabla^iR^{(3)}$ used below,
does not introduce the Ostrogradski instability (see, e.g., \cite{Gleyzes:2013ooa}).
According to the
Appendix. \ref{appA}, we have
\be
c_3= {aM_p^2\over
\gamma}\left(1+{2\tilde{m}_4^2\over M_p^2}\right),
\label{c3}
\ee
with $\gamma= H-(1/2)m^3_3/M_p^2$.
Again with Eq.(\ref{sigma1}), suppose $c_3\big{|}_{t_i}<0$, since
the integral $\int^{t_f}_{t_i}a dt$ diverges, eventually we
have $c_3\big{|}_{t_f}>0$, thus $c_3$ must cross $0$ at sometime $t$ with
$t_i<t<t_f$. However, if at that time we have ${2\tilde{m}_4^2/ M_p^2}$ cross $-1$,
the $c_3$ will cross $0$ naturally without the divergence of $\gamma$.
So the no-go behavior can be avoided, and notice for
Eq.(\ref{k2-correction1}), since $m_4^2(t)=0$ and $Q_T= 1$,
the tensor perturbation will be healthy.  Generally,
we could set the effective operator ${\tilde{m}_4^2}R^{(3)}\delta g^{00}/2$
to be dominated only when we meet $c_s^2<0$, thus it just modifies the sound
speed squared during this time, see Sec. \ref{Model} for details.

We can further add the term
$-m_4^2(t)\lf( \delta K^2-\delta K_{\mu\nu}\delta K^{\mu\nu} \rt)$
into the effective action Eq.(\ref{k2-correction1}), then $c_3$ changes to be
\be
c_3=
{a M_p^2 Q_T\over \gamma}\left(1+{2\tilde{m}_4^2\over
M_p^2}\right)
\label{mm}
\ee
with $\gamma=H\left(1+{2m_4^2\over
M_p^2}\right)-(1/2)m^3_3/M_p^2$.
Generally, we may find
those coefficients with ${2\tilde{m}_4^2/ M_p^2}$  crossing $-1$ and
$m_4^2(t)\neq \tilde{m}_4^2(t)$ which goes beyond Horndeski,
to make $c_3$ cross $0$ while $Q_T\neq 0$
and $\gamma$ won't blow up. The case of $Q_T/\gamma$ crosses $0$ is
discussed in \cite{Ijjas:2016vtq} with ${\cal L}_4$.

However, when $m_4^2(t)= \tilde{m}_4^2(t)$,
such as for the case of Horndeski theory,
\be
c_3={a M_p^2\over \gamma}Q_T^2
\ee
crosses $0$ suggests the no-go behavior must happen unless the integral
$\int^{t_f}_{t_i}adt$ in Eq.(\ref{k2-correction}) is convergent or
$Q_T^2/\gamma$ crosses $0$.
Obviously, this argument also applies to a general ${\cal L}_4$ with time-dependent $f(t)$,
as has been argued by T. Kobayashi \cite{Kobayashi:2016xpl} (see also \cite{Ijjas:2016vtq}).

Furthermore, let's consider the effective operators $(R^{(3)})^2$ and
$\nabla_iR^{(3)}\nabla^iR^{(3)}$, which will give contributions to higher
order spatial derivatives with $k^4\zeta_k^2$ and $k^6\zeta_k^2$.
As the operator $(R^{(3)})^2$ has been applied to the nonsingular
cosmology in \cite{Pirtskhalava:2014esa}, here we'd like to take the
following nonsingular model
\ba
\label{correction-k6}
S_{eff} &=&\int
d^4x\sqrt{-g}\Big[ {M_p^2\over2}R-\Lambda(t)-c(t)g^{00}
\nn\\
&\,&+{M_2^4(t)\over2}(\delta g^{00})^2-{m_3^3(t)\over2}\delta
K\delta g^{00} -{\tilde{\lambda}(t)\over
M_p^2}\nabla_iR^{(3)}\nabla^iR^{(3)}\Big]
\ea
with $f(t)=1$, and the coefficients $c(t)$, $\Lambda(t)$,
$M_2^4(t)$ and $m_3^3(t)$ are taken according to the EFT mapping
for the cubic Galileon ${\cal L}_2+{\cal L}_3$ \cite{Gleyzes:2013ooa}.
Then we have an effective sound speed squared with
\begin{eqnarray}
{c}_{s,eff}^2(k)=c_s^2+{32\tilde{\lambda}\over M_p^2  a^2 z^2}k^4\,,
\label{effective-cs2}
\end{eqnarray}
where $z=\sqrt{2a^2 c_1}$ and $c_s^2$ are given by
Eq.(\ref{cs2}).

From the equation of motion of $\zeta$ Eq.(\ref{MSEQ}), we see that if
these effective operators with higher order spatial derivatives have not
been included, we'll have a solution of
$\zeta\sim e^{-i\sqrt{c^2_s}k\Delta\tau}$ which indicates an exponential
growth when $c_s^2<0$. However, the growth turns out to be negligible for
the perturbation modes with $k\Delta\tau\ll 1$, and can be quite dangerous
for the modes with $k\Delta\tau\gg1$ \cite{Battarra:2014tga}\cite{Koehn:2015vvy}.
So we may specify $\tilde{\lambda}(t)$ to make
${c}_{s,eff}^2(k)\sim c_s^2$ for the modes with $k\Delta\tau\ll 1$,
while make ${c}_{s,eff}^2(k)$ modified to be positive for the modes with
$k\Delta\tau\gg 1$. Then such kind of exponential growth of $\zeta$
due to $c_s^2<0$ can be removed.

\section{Primordial perturbation spectrum within the framework of EFT}
\label{EFT-Perturbation}

In the last section we have presented how to evade the no-go theorem
within framework of EFT, i.e., by adding the effective operators such as
$R^{(3)}\delta g^{00}$, $(R^{(3)})^2$, $\nabla_iR^{(3)}\nabla^iR^{(3)}$,
to the original nonsingular cosmological models based on the cubic Galileon.
One might ask if the perturbation spectrum will be modified due to these
operators. In this section, we study the perturbation evolution in detail,
and show that the predictions can be consistent with the observations.

The equation of motion of $\zeta$ is
\ba
u''+\lf({c}_{s,eff}^2(k)k^2-{z''\over z} \rt)u=0\,,
\label{MSEQ}
\ea
where
\begin{eqnarray}
{c}_{s,eff}^2(k)=c_s^2-{2 c_4\over
z^2}k^2+{32\tilde{\lambda}\over M_p^2 a^2 z^2}k^4
\label{barc}
\end{eqnarray}
with $u=z\zeta$, $z=\sqrt{2a^2 c_1}$ and $c_s^2$ is given by
Eq.(\ref{cs2}), the prime denotes the derivative with respect to the
conformal time $\tau=\int dt/a$.

To study the evolution of the primordial perturbation concretely,
let's consider a bounce inflation background in this section.  We can define
the ``bouncing phase" as the time interval during which the NEC is violated,
i.e., ${\dot H}>0$.
At the beginning time $\tau_{B-}$  and the ending time $\tau_{B+}$ of
the ``bouncing phase"  we have ${\dot H}=0$, while before the beginning
time and after the ending time, the NEC is restored and thus leads to
${\dot H}<0$.

By adding the effective operators like $R^{(3)}\delta g^{00}$,
$(R^{(3)})^2$ or $\nabla_iR^{(3)}\nabla^iR^{(3)}$ to the original
G-bounce models \cite{Qiu:2011cy}\cite{Easson:2011zy}, within the
framework of EFT the whole Lagrangian tends to be like the ones in
Eq.(\ref{k2-correction1}) or Eq.(\ref{correction-k6}). To cure the
gradient instability problem in these models, we can set the
corresponding operators to be dominated only during the duration
$\Delta\tau=\tau_{B+}-\tau_{B-}$ of the bouncing phase.
Thus Eq.(\ref{MSEQ})  can be written as
\ba
u''+\left(k^2-{z''\over z}
\right)u=0\,,\quad  (\tau<\tau_{B-}, \,\,\tau>\tau_{B+}),
\label{conexp}
\ea
\ba
u''+A^{2n} k^{2n}u=0\,,\quad
(\tau_{B-}<\tau<\tau_{B+}),
\label{bounce}
\ea
where $n\geq1$, and the corresponding operators will respectively contribute
$\sim k^2, k^4, k^6$ corrections to the equation of motion. To be rigorous, all
the coefficients $A^{2n}$ of the $k^{2n}$ terms should be time-dependent.
However, here we set $A^{2n}$ constant for simplicity.

During the contracting phase $\tau<\tau_{B-}$, the background can
be parameterized as
\ba
a_c=a_{B-}\lf({\tau-\tilde{\tau}_{B-}\over
\tau_{B-}-\tilde{\tau}_{B-}}\rt)^{1\over \epsilon_c-1}\,,
\ea
where $\tilde{\tau}_{B-}=\tau_{B-}-[(\epsilon_c-1){\cal H}_{B-}]^{-1}$,
and $\epsilon_c>3$ is a constant. Thus we have
\ba
{a_c''\over
a_c}={\nu_c^2-{1\over4}\over(\tau-\tilde{\tau}_{B-})^2}\,,
\ea
where $\nu_c={1/2}-{1\over \epsilon_c-1}$.  The solution of
Eq.(\ref{conexp}) can be given as
\ba
\label{u-contracting}
u_c={\sqrt{\pi}\over 2}\sqrt{|\tau-\tilde{\tau}_{B-}|} \Big[
c_{1,1}H^{(1)}_{\nu_c}(k|\tau-\tilde{\tau}_{B-}|)
+c_{1,2}H^{(2)}_{\nu_c}(k|\tau-\tilde{\tau}_{B-}|) \Big]\,,
\ea
where $H^{(1)}_{\nu}$ and $H^{(2)}_{\nu}$ are the $\nu$-th order
Hankel function of the first and the second kind.

Initially, the perturbations are deep inside the horizon. The
initial condition can be taken as $u\sim{1\over \sqrt{2k}}e^{-ik\tau}$,
thus
\be
\label{initial-contracting}
c_{1,1}=1\,,\quad c_{1,2}=0\,.
\ee

During the bouncing phase $\tau_{B-}<\tau<\tau_{B+}$, the solution
of Eq.(\ref{bounce}) is
\ba
\label{u-bouncing}
u_b=c_{2,1}\cdot
e^{iA^nk^n(\tau-\tau_B)}+c_{2,2}\cdot e^{-iA^nk^n(\tau-\tau_B)}\,,
\ea
where $c_{2,1}$ and $c_{2,2}$ are determined by the evolution
of the contracting phase. By considering the effective operators,
the effective sound speed squared  $c_{s,eff}^2>0$ for short wavelength
perturbation modes, thus there won't be any dangerous growths of the
curvature perturbation $\zeta$.

During the inflation $\tau>\tau_{B+}$, the background can be
parameterized as
\ba
a_e=a_{B+}\lf({\tau-\tilde{\tau}_{B+}\over
\tau_{B+}-\tilde{\tau}_{B+}}\rt)^{1\over \epsilon_e-1}\,,
\ea
where $\tilde{\tau}_{B+}=\tau_{B+}-[(\epsilon_e-1){\cal
H}_{B+}]^{-1}$. So we have
\ba
{a_e''\over
a_e}={\nu_e^2-{1\over4}\over(\tau-\tilde{\tau}_{B+})^2}\,,
\ea
where $\nu_e={1/2}-{1\over \epsilon_e-1}$. The solution of
Eq.(\ref{conexp}) can be given as
\ba
\label{u-expanding}
u_e={\sqrt{\pi}\over 2}\sqrt{|\tau-\tilde{\tau}_{B+}|} \Big[
c_{3,1}H^{(1)}_{\nu_e}(k|\tau-\tilde{\tau}_{B+}|)
+c_{3,2}H^{(2)}_{\nu_e}(k|\tau-\tilde{\tau}_{B+}|) \Big]\,.
\ea

The power spectrum is calculated as
\ba
P_{\zeta}=P_{\zeta}^{inf} \cdot|c_{3,1}-c_{3,2}|^2\,.
\ea
The information of the evolution
history of the universe and the contributions of the EFT operators
are encoded in $c_{3,1}$ and $c_{3,2}$. Though we work with
bounce inflation scenario, actually, our result is also applicable
to the bounce scenario, as will be seen.

By requiring the continuity of $u$ and $u'$ at the matching
surfaces, we obtain
\ba
\label{bb}
\left(
\begin{array}{ccc} c_{3,1}\\c_{3,2}
\end{array}\right)&=&{\cal
M}^{(3,2)}\times{\cal
M}^{(2,1)}\times\left(\begin{array}{ccc}c_{1, 1}\\c_{1,
2}\end{array}\right), \label{rec1}
\ea
where the components of the matrix ${\cal M}^{(2,1)}$ are
\ba
{\cal M}_{11}^{(2,1)}&=& {e^{i d
A^nk^n}\sqrt{\pi}\over 8A^nk^n \sqrt{\hat{\cal H}}} \Big[ ik
H^{(1)}_{\nu_c-1}\lf({k\over \hat {\cal H}}\rt) -ik
H^{(1)}_{\nu_c+1}\lf({k\over \hat {\cal H}}\rt) +(2A^nk^n+i \hat
{\cal H})H^{(1)}_{\nu_c}\lf({k\over \hat {\cal H}}\rt) \Big] ,
\nonumber\\
{\cal M}_{12}^{(2,1)}&=& {e^{i d A^nk^n}\sqrt{\pi}\over 8A^nk^n
\sqrt{\hat{{\cal H}}}} \Big[ ik H^{(2)}_{\nu_c-1}\lf({k\over \hat
{\cal H}}\rt) -ik H^{(2)}_{\nu_c+1}\lf({k\over \hat {\cal H}}\rt)
+(2A^nk^n+i \hat {\cal H})H^{(2)}_{\nu_c}\lf({k\over \hat {\cal
H}}\rt) \Big] ,
\nonumber\\
{\cal M}_{21}^{(2,1)}&=& {e^{-i d A^nk^n}\sqrt{\pi}\over 8A^nk^n
\sqrt{\hat{{\cal H}}}} \Big[ -ik H^{(1)}_{\nu_c-1}\lf({k\over \hat
{\cal H}}\rt) +ik H^{(1)}_{\nu_c+1}\lf({k\over \hat {\cal H}}\rt)
+(2A^nk^n-i \hat {\cal H})H^{(1)}_{\nu_c}\lf({k\over \hat {\cal
H}}\rt) \Big] ,
\nonumber\\
{\cal M}_{22}^{(2,1)}&=& {e^{-i d A^nk^n}\sqrt{\pi}\over 8A^nk^n
\sqrt{\hat{{\cal H}}}} \Big[ -ik H^{(2)}_{\nu_c-1}\lf({k\over \hat
{\cal H}}\rt) +ik H^{(2)}_{\nu_c+1}\lf({k\over \hat {\cal H}}\rt)
+(2A^nk^n-i \hat {\cal H})H^{(2)}_{\nu_c}\lf({k\over \hat {\cal
H}}\rt) \Big] ,\nonumber\label{matrix1}
\ea
and the components of matrix ${\cal M}^{(3,2)}$ are
\ba
{\cal
M}_{11}^{(3,2)}&=& -{i e^{idA^nk^n}\sqrt{\pi}\over 4\sqrt{ {\cal
H}_{B+}}} \Big[ -2k H^{(2)}_{\nu_e-1}\lf({k\over {\cal H}_{B+} }
\rt) +\Big( -2iA^nk^n+(2\nu_e-1){\cal H}_{B+} \Big)
H^{(2)}_{\nu_e}\lf({k\over {\cal H}_{B+} } \rt) \Big] ,
\nonumber\\
{\cal M}_{12}^{(3,2)}&=& { e^{-idA^nk^n}\sqrt{\pi}\over 4\sqrt{
{\cal H}_{B+}}} \Big[ 2ik H^{(2)}_{\nu_e-1}\lf({k\over {\cal
H}_{B+} } \rt) +\Big( 2A^nk^n-i(2\nu_e-1){\cal H}_{B+} \Big)
H^{(2)}_{\nu_e}\lf({k\over {\cal H}_{B+} } \rt) \Big] ,
\nonumber\\
{\cal M}_{21}^{(3,2)}&=& { e^{idA^nk^n}\sqrt{\pi}\over 4\sqrt{
{\cal H}_{B+}}} \Big[ -2ik H^{(1)}_{\nu_e-1}\lf({k\over {\cal
H}_{B+} } \rt) +\Big( 2A^nk^n+i(2\nu_e-1){\cal H}_{B+} \Big)
H^{(1)}_{\nu_e}\lf({k\over {\cal H}_{B+} } \rt) \Big] ,
\nonumber\\
{\cal M}_{22}^{(3,2)}&=& {i e^{-idA^nk^n}\sqrt{\pi}\over 4\sqrt{
{\cal H}_{B+}}} \Big[ -2k H^{(1)}_{\nu_e-1}\lf({k\over {\cal
H}_{B+} } \rt) +\Big( 2iA^nk^n+(2\nu_e-1){\cal H}_{B+} \Big)
H^{(1)}_{\nu_e}\lf({k\over {\cal H}_{B+} } \rt) \Big]
\label{matrix2} \nonumber
\ea
with $d=\tau_{B+}-\tau_B$, $\hat{\cal H}=(\epsilon_c-1) {\cal H}_{B+}$,
and ${\cal H}_{B+}$ is the comoving Hubble parameter at $\tau_{B+}$.

Considering the long wavelength limit, $k/{\cal H}_{B+}\ll1$, we have
\ba
|c_{3,1}-c_{3,2}|^2 &\approx& {1\over 9\pi}\lf( {k\over {\cal
H}_{B+} } \rt)^{2\epsilon_c\over \epsilon_c-1} (1-4d{\cal
H}_{B+})^2(2\epsilon_c-2)^{2\over 1-\epsilon_c} \Gamma^2\lf(
{1\over2}+{1\over 1-\epsilon_c} \rt)
\nn\\
&\sim& \lf( {k\over {\cal H}_{B+} } \rt)^{2\epsilon_c\over
\epsilon_c-1}\,. \label{smallk}
\ea

In bounce scenario where the bounce is followed by the Hot Big-Bang
expansion, $P_{\zeta}$ is given by Eq.(\ref{smallk}),
\ba
P_{\zeta}\sim
|c_{3,1}-c_{3,2}|^2 \sim \lf( {k\over {\cal H}_{B+} }
\rt)^{2\epsilon_c\over \epsilon_c-1}\,
\ea
since the perturbation modes with $k/{\cal H}_{B+}>1$ can be
hardly produced during the expansion after the bounce. The result is
consistent with that in ekpyrotic universe\cite{Khoury:2001wf}
\cite{Lehners:2007ac}\cite{Buchbinder:2007ad}. Thus the spectrum
of primordial perturbations in bounce scenario is unaffected by the
corresponding spatial derivative operators in Eq.(\ref{eft_action}).

However, in the bounce inflation scenario, the perturbation modes with
$k/{\cal H}_{B+}>1$ will be produced during the inflation after the bounce.
When we take the short wavelength limit, $k/{\cal H}_{B+}\gg 1$, the
$|c_{3,1}-c_{3,2}|^2$ acquires drastic oscillation and even
diverges when $k/{\cal H}_{B+}\rightarrow \infty$. Without making
qualitative deviation, we have
\ba
\label{amplification}
|c_{3,1}-c_{3,2}|^2 &\approx& 1+\lf( {k\over {\cal H}_{B+} }
\rt)^{2n-2}{A^{2n}\over {\cal H}_{B+}^{2-2n} } \cos^2\lf( {k\over
{\cal H}_{B+} } \rt) \sin^2(2dA^nk^n)\,.\label{largek}
\ea
The
``1" in right-hand side of Eq.(\ref{largek}) actually stands for
the terms $\sim k^0$, such as $\cos^2(2dA^nk^n)$. Here, we do not
specify it, since it makes no qualitative difference when $k/{\cal
H}_{B+}\gg 1$. In order to satisfy the observations, Eq.(\ref{largek})
should be nearly scale invariant. Thus the operator $R^{(3)}\delta
g^{00}\sim k^2$ in EFT Eq.(\ref{eft_action}) is applicable, but the
operators $(R^{(3)})^2\sim k^4$ and
$\nabla_iR^{(3)}\nabla^iR^{(3)}\sim k^6$ will make Eq.(\ref{largek})
diverge, since
\be
|c_{3,1}-c_{3,2}|^2\sim \left({k\over {\cal
H}_{B+}}\right)^{2n-2},
\ee
for $k/{\cal H}_{B+}\gg 1$, which are
unacceptable. This result could be general, though the drastic
oscillations in Eq.(\ref{largek}) might be attributed to the
matching method and the oversimplified approximation we have
used.

\section{Application: constructing a healthy G-bounce inflation model}
\label{Model}

In this section we apply the effective operator $R^{(3)}\delta g^{00}$
to cure the gradient instability faced by the G-bounce inflation model
proposed in \cite{Qiu:2015nha} (see also \cite{Wan:2015hya}).  The
G-bounce inflation background was built by using the cubic Galileon,
which can be written in the EFT language as
\begin{eqnarray}
\label{G-bounce}
c(t) &=&\frac{1}{2}\dot{\phi}^2_0\left({\cal K}(\phi)+{\cal
T}\dot{\phi}^2_0\right)
+\frac{1}{2}\dot{\phi}^2_0\left(-\ddot{\phi}_0+3H\dot{\phi}_0\right)
G_{3X}-\dot{\phi}^2_0G_{3\phi}, \nonumber\\ \Lambda(t)
&=&\frac{1}{4}{\cal T}(\phi)\dot{\phi}^4_0+V(\phi)+\frac{1}{2}
\dot{\phi}^2_0\left(\ddot{\phi}_0+3H\dot{\phi}\right)G_{3X}, \nonumber\\
\nonumber M^4_2 &=&\frac{1}{2}{\cal
T}(\phi)\dot{\phi}^4_0+\frac{1}{4}\left(\ddot{\phi}_0+
3H\dot{\phi}_0\right)\dot{\phi}^2_0G_{3X} +\frac{3}{4}H\dot{\phi}^5_0
G_{3XX}-\frac{1}{4}\dot{\phi}^4_0 G_{3X\phi}, \nonumber\\
m^3_3&=&\dot{\phi}^3_0 G_{3X},
\end{eqnarray}
where \ba
 \label{KTG-GBinf}
&&{\cal
K}(\phi)=1-2k_0\left[1+2\kappa_1\left(\frac{\phi}{M_p}\right)^2\right]^{-2}~,~~~
{\cal T}(\phi)=\frac{t_0}{M_p^4}\left[1+2\kappa_2\left(\frac{\phi}{M_p}\right)^2\right]^{-2}~,\nonumber\\
&&G_3\left(\phi,X\right)=\frac{\theta
X}{M_p^3}\left[1+2\kappa_2\left(\frac{\phi}{M_p}\right)^2\right]^{-2}~,
\ea
and
\begin{eqnarray}
\label{vphi}
&&V(\phi)=-V_0 e^{\bar{c}\phi/M_p}\left[1-\tanh(\lambda_1
\frac{\phi}{M_p})\right] +\Lambda^4_{inf}
\left(1-\frac{\phi^2}{v^2}\right)^2 \left[1+\tanh(\lambda_2
\frac{\phi}{M_p})\right]
\end{eqnarray}
such that $ V=-V_0 e^{\bar{c}\phi}$ for $\phi\ll -M_p/\lambda_1$
(responsible for the ekpyrotic contraction), and is
$V=\Lambda^4_{inf}(1-\frac{\phi^2}{v^2})^2$ for $\phi\gg
M_p/\lambda_2$ (responsible for the inflation after bounce).
Here $k_0$, $t_0$, $\theta$, $\kappa_1$, $\kappa_2$, $\lambda_1$,
$\lambda_2$, $V_0$, $\bar{c}$, $\Lambda_{inf}$ and $v$ are constants.

However, the bounce with the cubic Galileon is pathological due to the
existence of the no-go theorem. Actually, the gradient instability exists
since $c_s^2<0$ around the bounce \cite{Qiu:2015nha}\cite{Wan:2015hya}.
As has been argued in Sec.\ref{evadenogo}, it can be avoided by introducing
an effective  operator ${{\tilde m}_4^2\over 2}R^{(3)}\delta g^{00}$.
By doing so, $c_s^2$ is modified to
\be
c_s^2={c_3'-a^2M_p^2\over a^2 c_1}\,,
\ee
where
\ba
c_1&=&{M_p^2\over (2HM_p^2-m_3^3)^2}\lf( 3m_3^6+4H^2\epsilon M_p^4+8M_p^2M_2^4 \rt)\,,
\\
c_3&=&{aM_p^2\over H-{m_3^3/(2 M_p^2)} }\lf( 1+2{\tilde{m}_4^2\over M_p^2 } \rt)\,.
\ea

\begin{figure}[htbp]
\subfigure[~~$c_s^2$]{\includegraphics[width=.48\textwidth]{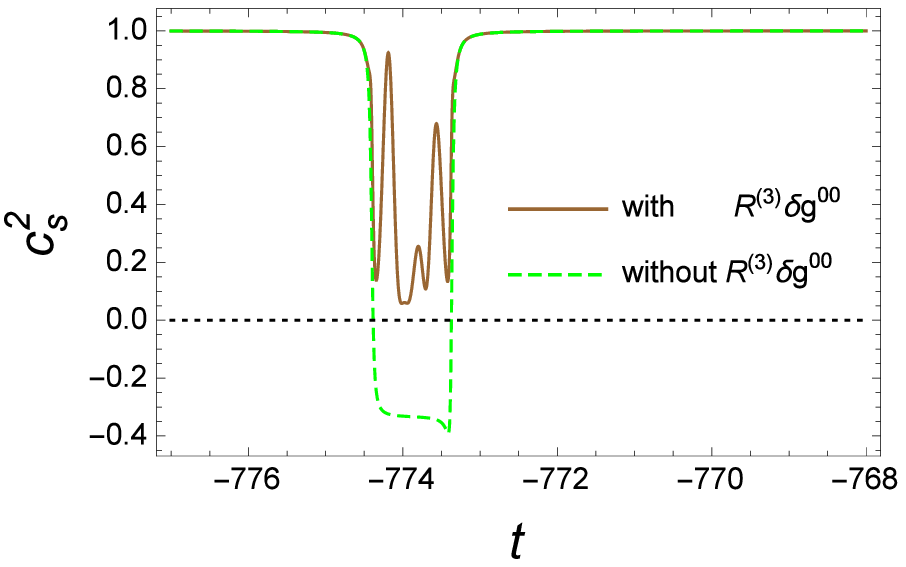} }
\subfigure[~~$\tilde m _4^2/M_p^2$]{\includegraphics[width=.48\textwidth]{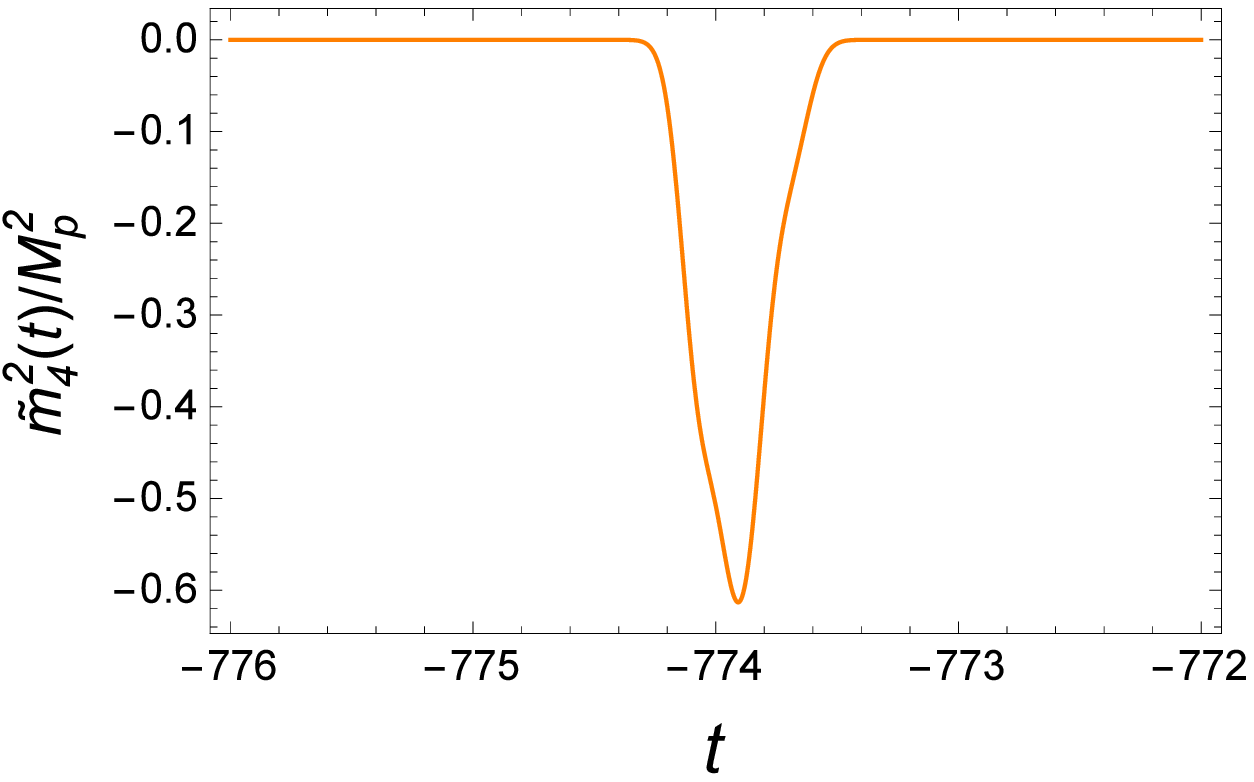} }
\caption{Left: the evolution of $c_s^2$ in G-bounce inflation model
\cite{Qiu:2015nha}, right: the function of $\tilde m _4^2$. We can see
the $c_s^2$ can be modified to be larger than 0 by introducing the
effective operator ${{\tilde m}_4^2\over 2}R^{(3)}\delta g^{00}$.}
\label{shapefunction}
\end{figure}

We are able to avoid the gradient instability by choosing
a suitable ${\tilde m}_4^2(t)$. We have numerically calculated
Eqs.(\ref{G-bounce}), see e.g., \cite{Qiu:2015nha}\cite{Li:2016awk},
and plotted the evolution of $c_s^2$ in Fig. \ref{shapefunction}.
The effect of $R^{(3)}\delta g^{00}$ on $c_s^2$ can be clearly seen
in  Fig. \ref{shapefunction}. Because $c_1$ is unaffected by
$\tilde m _4^2$, there is also no ghost instability, as demonstrated
in \cite{Qiu:2015nha}. Noting that the operator $\xi(t) R^{(3)}$ used
in Ref.\cite{Qiu:2015nha} dose not involve $R^{(3)}\delta g^{00}$.

\section{Conclusion}
\label{Conclusion}

Building classical nonsingular cosmological models is inspiring, since it
offers us a self-consistent framework to deeply understand the
physics of the primordial universe, even though we still don't know
the complete theory of the quantum gravity. However, the popular
nonsingular cosmological models based on the cubic Galileon are
afflicted by the LMR no-go theorem, which means we have to go
beyond the cubic Galileon to construct models without pathologies.

In this paper, we have explored the nonsingular cosmology within the
framework of EFT. We have illustrated how to avoid the no-go theorem
in theories beyond Horndeski, and pointed out how could the effective
operators, such as $R^{(3)}\delta g^{00}$, $(R^{(3)})^2$ and
$\nabla_iR^{(3)}\nabla^iR^{(3)}$, play significant roles in
building healthy nonsingular cosmological models. We also have
studied the perturbation evolution of these healthy models.
We find that the spectrum of the primordial perturbation can be
consistent with the observations.

We conclude that based on EFT, a fully healthy nonsingular bounce
model can be built without begging any unknown physics. As an
application of the EFT, we have presented a realistic healthy bounce
inflation model by making use of the operator $R^{(3)}\delta g^{00}$.
The study of classical nonsingular cosmology in the framework of EFT will
be helpful for understanding the evolution and the gravity theory in the
primordial universe.

\begin{acknowledgments}
We thank Xinmin Zhang and Yi-Fu Cai for helpful discussions.
YC thanks Paul J. Steinhardt and Anna Ijjas for valuable
communications.
YW wishes to thank Y. Piao and School of Physics, University of
Chinese Academy of Sciences in
Beijing for hospitality during the time while this work was
started. YW also wishes to thank T. Qiu and Institute of
Astrophysics, Central China Normal University in Wuhan for
hospitality. YP is supported by NSFC, No. 11222546, 11575188, and
the Strategic Priority Research Program of Chinese Academy of
Sciences, No. XDB23010100. TQ is supported in part
by NSFC under Grant No. 11405069 and in part by the Open
Innovation Fund of Key Laboratory of Quark and Lepton
Physics (MOE), Central China Normal University
(No. QLPL2014P01).

\end{acknowledgments}

\appendix

\section{The derivations of the quadratic actions for scalar and tensor perturbations }
\label{appA}

With the ADM line element given in Eq.(\ref{metric}), we have
\begin{equation}
g_{\mu\nu}=\left(
  \begin{array}{cc}
  N_kN^k-N^2 &  N_j\\
  N_i &  h_{ij}\\
  \end{array}
\right) \,,\qquad
g^{\mu\nu}=\left(
  \begin{array}{cc}
  -N^{-2} &  {N^j\over N^2}\\
  {N^i\over N^2} &  h^{ij}-{N^iN^j\over N^2}\\
  \end{array}
\right) \,,\qquad
\end{equation}
and $\sqrt{-g}=N\sqrt{h}$,
where $N_i=h_{ij}N^j$, and the spatial indices are raised
and lowered by the spatial metric $h_{ij}$.
We can define the unit one-form tangent vector
$n_{\nu}=n_0 (dt/dx^{\mu})=(-N,0,0,0)$ and $n^{\nu}=g^{\mu\nu}n_{\mu} =({1/ N},-{N^i/ N})$,
which satisfies $n_{\mu}n^{\mu}=-1$.
The induced 3-dimensional metric on the  hypersurface is
$H_{\mu\nu}=g_{\mu\nu}+n_{\mu}n_{\nu}$, thus
\begin{equation}
H_{\mu\nu}=\left(
  \begin{array}{cc}
  N_kN^k &  N_j\\
  N_i &  h_{ij}\\
  \end{array}
\right) \,,\qquad
H^{\mu\nu}=\left(
  \begin{array}{cc}
  0 &  0\\
  0 &  h^{ij}\\
  \end{array}
\right) \,.\qquad
\end{equation}

Moreover, the extrinsic curvature on the  hypersurface is
\ba K_{\mu\nu}&\equiv&{1\over2}{\cal L}_{n}H_{\mu\nu}\nn\\
&=&{1\over 2N}(\dot{H}_{\mu\nu}-D_{\mu}N_{\nu}-D_{\nu}N_{\mu})~,
\ea
where ${\cal L}_{n}$  is the Lie derivative with respective to $n^{\mu}$, and $D_{\mu}$ is the covariant derivative associate with $H_{\mu\nu}$. The Ricci scalar is decomposed as
\be
\label{Ricci}
R=R^{(3)}-K^2+K_{\mu\nu}K^{\mu\nu}+2\nabla_\mu(Kn^\mu-n^\nu\nabla_\nu n^\mu)~.
\ee
where $R^{(3)}$ is the induced 3-dimensional Ricci scalar associated with $H_{\mu\nu}$.
Note that in general, when there is a non-minimal coupling between the scalar field and $R$,
the last term in Eq.(\ref{Ricci}) cannot be discarded.

In action (\ref{eft_action}), we have defined
\ba
&\,&\delta g^{00}=g^{00}+1\,,\\
&\,&\delta K_{\mu\nu}=K_{\mu\nu}-H_{\mu\nu}H\,,\\
&\,&\delta K^{\mu\nu}=K^{\mu\nu}-H^{\mu\nu}H\,,\\
&\,&\delta K=\delta K^{\mu}_{\mu}=K^{\mu}_{\mu}-3H\,.
\ea

In the unitary gauge, we set
\be h_{ij}=a^2e^{2\zeta}(e^{\gamma})_{ij},\qquad \gamma_{ii}=0=\partial_i\gamma_{ij} \,.
\ee
Moreover, $N$ and $N_i$ are expressed as $N=1+\alpha$ and $N_i=\partial_i\beta$.
Then, it is straightforward to obtain
\ba
\label{dg00}
&\,&\delta g^{00}=1-{1\over(1+\alpha)^2}\,,\\
\label{r3}
&\,& R^{(3)}=-2a^{-2}e^{-2\zeta}\Big[ 2\partial^2\zeta+(\partial\zeta)^2 \Big]\,,
\\
\label{dkxij}
&\,& \delta K_{ij}
={1\over 1+\alpha}\lf\{a^2(\dot{\zeta}-\alpha H)e^{2\zeta}\delta_{ij}-\partial_i\partial_j\beta
+\partial_i\beta\partial_j\zeta+\partial_j\beta\partial_i\zeta
-\partial_k\beta\partial_k\zeta\delta_{ij}  \rt\}\,,
\\
\label{dksij}
&\,& \delta K^{ij}
=
{ a^{-4}e^{-4\zeta} \over 1+\alpha }\lf\{a^2(\dot{\zeta}-\alpha H)e^{2\zeta}\delta^{ij}-\partial_i\partial_j\beta
+\partial_i\beta\partial_j\zeta+\partial_j\beta\partial_i\zeta
-\partial_k\beta\partial_k\zeta\delta_{ij}  \rt\}\,,
\ea
where $\partial^2=\partial_i\partial_i$.

Substituting Eqs.(\ref{dg00}) to (\ref{dksij}) into the action (\ref{eft_action}) and using
the Hamiltonian constraints
\be
{\partial {\cal L}\over \partial \alpha}=0\,, \qquad {\partial {\cal L}\over \partial (\partial^2\beta)}=0\,,
\ee
we find
\be
\alpha=A_1 \dot{\zeta}+A_2\partial^2\zeta\,,
\qquad
\partial^2\beta=B_1 \dot{\zeta}+B_2\partial^2\zeta\,,\label{constr}
\ee
in which
\ba
A_1&=&{2\over D}{ \left(f M_p^2
 +2 m_ 4^2 \right)
 \left[
  \left(2 f H
 + \dot f\right) M_p^2
 - m_3^3
 + 4 H m_ 4^2
 + 6 H \bar{m}_ 4^2\right]}\,,
 \nn\\
 A_2&=&
 \frac{2}{a^2 D}\lf\{
 M_p^2 \left[2 f \bar{m}_ 4^2
 + \left(2 f H
 + \dot f\right) \bar{m}_ 5\right]
 - \left(m_ 3^3
 - 4 H m_ 4^2\right) \bar{m}_ 5
 + 4 \bar{m}_ 4^2 \tilde{m}_ 4^2
 \rt\}\,,
 \nn\\
 B_1&=&
 {a^2\over {D}}
 \Big\{
  3 m_ 3^6
 - 6 \dot f m_ 3^3 M_p^2
 + 8 f M_ 2^4 M_p^2
 + \left(4 f^2 H^2 \epsilon
 + 2 f \dot{f} H
 + 3 \dot f{}^2
 - 2 f \ddot f\right) M_p^4
 \nn\\&\,&
 + (4 m_ 4^2+6\bar{m}_ 4^2) \left[4 M_ 2^4
 +  ( 2 f H^2 \epsilon
 + \dot f H
 - \ddot f ) M_p^2\right]
 \Big\}
 \,,
 \nn\\
 B_2&=&
 {2\over D}
 \Big\{ \left[3 H m_ 3^3
 + 4 M_ 2^4
 + \left(2 f H^2 \epsilon
 - 2 H \dot f
 - \ddot f\right) M_p^2\right] \bar{m}_ 5
 \nn\\&\,&
 - \left[
 \left(2 f H
 + \dot f\right) M_p^2
 - m_3^3
 + 4 H m_ 4^2
 + 6 H \bar{m}_ 4^2\right] \left(
   f M_p^2
 + 2 \tilde{m}_ 4^2\right)\Big\}\,,
 \nn\\
 D&=&
 \left[m_ 3^3
 - 4 H m_ 4^2
 - \left(2 f H
 + \dot f\right) M_p^2\right]^2
 \nn\\&\,&
 + 2\bar{m}_ 4^2 \left[12 H^2 m_ 4^2
 + 4 M_ 2^4
 + \left(   f H^2 (6
 + 2\epsilon )
 + \dot f H
 - \ddot f\right) M_p^2\right]\,.
\ea

Then, with Eqs.(\ref{constr}), we obtain the quadratic action of scalar
perturbation, which is displayed in Eq.(\ref{eft_action02}). Here, we
write down the expressions of the coefficients in Eq.(\ref{eft_action02}):

\ba
c_1&=&
\frac{1}{D}
\left(2 m_ 4^2
 + f M_p^2\right) \Big\{3 m_ 3^6
 + 4 f^2 H^2 \epsilon  M_p^4
 + 8 M_ 2^4 \left(2 m_ 4^2
 + 3 \bar{m}_ 4^2\right)
 \nn\\&\,&
 + M_p^2 \left[
 - 2 \ddot f \left(2 m_ 4^2
 + 3 \bar{m}_ 4^2\right)
 + \dot f \left(
 - 6 m_ 3^3
 + 4 H m_ 4^2
 + 3 \dot f M_p^2
 + 6 H \bar{m}_ 4^2\right)\right]
 \nn\\&\,&
 + 2 f M_p^2 \left[4 M_ 2^4
 - \ddot f M_p^2
 + H \left(4 H \epsilon  m_ 4^2
 + \dot f M_p^2
 + 6 H \epsilon  \bar{m}_ 4^2\right)\right]\Big\}  \,,
 \nn\\
 c_2&=&f M_p^2\,,
 \nn\\
 c_3&=&
 \frac{2 a}{D}
  \left(2 m_ 4^2
 + f M_p^2\right)
 \Big\{2 f^2 H M_p^4
 + \bar{m}_ 5 \lf[(2 H \dot f+ \ddot f) M_p^2
 - 3 H m_ 3^3
 - 4 M_ 2^4\rt]
 \nn\\&\,&
 + f M_p^2 \left[
 - m_3^3
 + \dot f M_p^2
 + 2 H \left(2 m_ 4^2
 + 3 \bar{m}_ 4^2
 - H \epsilon  \bar{m}_ 5
 + 2 \tilde{m}_ 4^2\right)\right]
  \nn\\&\,&
 +\tilde{m}_ 4^2 \lf( 8 H m_ 4^2
 - 2 m_ 3^3
 + 2 \dot f M_p^2
 + 12 H \bar{m}_ 4^2   \rt)
 \Big\}\,,
 \nn\\
c_4
 &=&{2\over D}\Big\{
 4 \bar \lambda D
 +
 \left[12 H^2 m_ 4^2
 + 4 M_ 2^4
 + \left(H \dot f
 - \ddot f\right) M_p^2\right] \bar{m}_ 5^2
  \nn\\&\,&
 -2 f^2 M_p^4 \left(\bar{m}_ 4^2
 + 2 H \bar{m}_ 5\right)
 +4 \left(m_ 3^3
 - 4 H m_ 4^2
 - \dot f M_p^2\right) \bar{m}_ 5 \tilde{m}_ 4^2
 \nn\\&\,&
 +2 f M_p^2 \bar{m}_ 5 \left[m_ 3^3
 - 4 H m_ 4^2
 - \dot f M_p^2
 + H^2 (3
 + \epsilon ) \bar{m}_ 5\right]
 \nn\\&\,&
 - 8 f M_p^2 \left(\bar{m}_ 4^2
 + H \bar{m}_ 5\right) \tilde{m}_ 4^2
 - 8 \bar{m}_ 4^2 \tilde{m}_ 4^4
 \Big\}
 \,.
\ea

As for the tensorial part, we have $N=1$, $N_i=0$ and $\zeta=0$. It is also straightforward to obtain
\ba
&\,& R^{(3)}=-{1\over4}a^{-2}\gamma^{kl,i}\gamma_{kl,i}+{\cal O}(\gamma^3)\,,
\\
&\,& K_{ij}=
a^2\Big[H\delta_{ij}+H\gamma_{ij}+{1\over2}\dot{\gamma}_{ij}+
{1\over2}H\gamma_{ik}\gamma^k_j
+{1\over4}(\dot{\gamma}_{ik}\gamma^k_j+\gamma_{ik}\dot{\gamma}^k_j)
 \Big]+{\cal O}(\gamma^3)\,,
\\
&\,& { \delta K_{ij}}=
{a^2\over2}\Big[\dot{\gamma}_{ij}
+{1\over2}(\dot{\gamma}_{ik}\gamma^k_j+\gamma_{ik}\dot{\gamma}^k_j)\Big]+{\cal O}(\gamma^3)\,.
\\
&\,& K^{ij}=a^{-2}\Big[H\delta^{ij}+{1\over2}\dot{\gamma}^{ij}-H\gamma^{ij}
-{1\over4}(\dot{\gamma}^i_l\gamma^{lj}+\gamma^i_l\dot{\gamma}^{lj})
+{1\over2}H\gamma^{je}\gamma_e^i\Big]+{\cal O}(\gamma^3)\,,
\\
&\,& { \delta K^{ij}}={a^{-2}\over2}\Big[\dot{\gamma}^{ij}-{1\over2}
(\dot{\gamma}^i_l\gamma^{lj}+\gamma^i_l\dot{\gamma}^{lj})
\Big]+{\cal O}(\gamma^3)\,,
\\
&\,& K= 3H+{\cal O}(\gamma^3) \,.
\ea
Note that $\delta K=K-3H$ contains only scalars up to the quadratic order, as well as
the last term in Eq.(\ref{Ricci}). Substituting the above results into action (\ref{eft_action}), we obtain the quadratic action of tensor perturbation, which is displayed in Eq.(\ref{tensor-action}).


\begin{thebibliography}{99}



\bibitem{Borde:1993xh}
  A.~Borde and A.~Vilenkin,
  Phys.\ Rev.\ Lett.\  {\bf 72}, 3305 (1994)

\bibitem{Borde:2001nh}
  A.~Borde, A.~H.~Guth and A.~Vilenkin,
  Phys.\ Rev.\ Lett.\  {\bf 90}, 151301 (2003)


\bibitem{Piao:2003zm}
  Y.~-S.~Piao, B.~Feng and X.~-m.~Zhang,
  Phys.\ Rev.\ D {\bf 69}, 103520 (2004)
  [hep-th/0310206];
  Y.~-S.~Piao,
  Phys.\ Rev.\ D {\bf 71}, 087301 (2005)
  [astro-ph/0502343]; 
  Y.~-S.~Piao, S.~Tsujikawa and X.~-m.~Zhang,
  Class.\ Quant.\ Grav.\  {\bf 21}, 4455 (2004)  [hep-th/0312139].


\bibitem{Liu:2013kea}
  Z.~G.~Liu, Z.~K.~Guo and Y.~S.~Piao,
  Phys.\ Rev.\ D {\bf 88}, 063539 (2013)
  [arXiv:1304.6527 [astro-ph.CO]].

\bibitem{Qiu:2015nha}
  T.~Qiu and Y.~T.~Wang,
  JHEP {\bf 1504}, 130 (2015)
  [arXiv:1501.03568 [astro-ph.CO]].

\bibitem{Liu:2014tda}
  Z.~G.~Liu, H.~Li and Y.~S.~Piao,
  Phys.\ Rev.\ D {\bf 90}, no. 8, 083521 (2014)
  [arXiv:1405.1188 [astro-ph.CO]].

\bibitem{Pirtskhalava:2014esa}
  D.~Pirtskhalava, L.~Santoni, E.~Trincherini and P.~Uttayarat,
  JHEP {\bf 1412}, 151 (2014)
  [arXiv:1410.0882 [hep-th]].


\bibitem{Kobayashi:2015gga}
  T.~Kobayashi, M.~Yamaguchi and J.~Yokoyama,
  JCAP {\bf 1507}, no. 07, 017 (2015)
  [arXiv:1504.05710 [hep-th]].


\bibitem{Cai:2007qw}
  Y.~F.~Cai, T.~Qiu, Y.~S.~Piao, M.~Li and X.~Zhang,
  JHEP {\bf 0710}, 071 (2007)
  [arXiv:0704.1090 [gr-qc]].

\bibitem{Cai:2008qw}
  Y.~F.~Cai, T.~t.~Qiu, R.~Brandenberger and X.~m.~Zhang,
  Phys.\ Rev.\ D {\bf 80}, 023511 (2009)
  [arXiv:0810.4677 [hep-th]].


\bibitem{Creminelli:2010ba}
  P.~Creminelli, A.~Nicolis and E.~Trincherini,
  JCAP {\bf 1011}, 021 (2010)
  [arXiv:1007.0027 [hep-th]].



\bibitem{Hinterbichler:2012yn}
  K.~Hinterbichler, A.~Joyce, J.~Khoury and G.~E.~J.~Miller,
  Phys.\ Rev.\ Lett.\  {\bf 110}, 24, 241303 (2013)
  [arXiv:1212.3607 [hep-th]].



\bibitem{Cai:2013rna}
  Y.~F.~Cai, Y.~Wan and X.~Zhang,
  Phys.\ Lett.\ B {\bf 731}, 217 (2014)
  [arXiv:1312.0740 [hep-th]].

\bibitem{Piao:2003ty}
  Y.~S.~Piao and E.~Zhou,
  Phys.\ Rev.\ D {\bf 68}, 083515 (2003)
  [hep-th/0308080].


\bibitem{Liu:2011ns}
  Z.~G.~Liu, J.~Zhang and Y.~S.~Piao,
  Phys.\ Rev.\ D {\bf 84}, 063508 (2011)  [arXiv:1105.5713];
  Y.~Cai and Y.~S.~Piao,
  arXiv:1601.07031 [hep-th].



\bibitem{Liu:2012ww}
  Z.~G.~Liu and Y.~S.~Piao,
  Phys.\ Lett.\ B {\bf 718}, 734 (2013)
  [arXiv:1207.2568 [gr-qc]].


\bibitem{Rubakov:2014jja}
  V.~A.~Rubakov,
  Phys.\ Usp.\  {\bf 57}, 128 (2014)
  [Usp.\ Fiz.\ Nauk {\bf 184}, no. 2, 137 (2014)]
  [arXiv:1401.4024 [hep-th]].

\bibitem{Carroll:2003st}
  S.~M.~Carroll, M.~Hoffman and M.~Trodden,
  Phys.\ Rev.\ D {\bf 68}, 023509 (2003)
  [astro-ph/0301273];
  J.~M.~Cline, S.~Jeon and G.~D.~Moore,
  Phys.\ Rev.\ D {\bf 70}, 043543 (2004)
  [hep-ph/0311312].

\bibitem{Nicolis:2008in}
  A.~Nicolis, R.~Rattazzi and E.~Trincherini,
  Phys.\ Rev.\ D {\bf 79}, 064036 (2009)
  [arXiv:0811.2197 [hep-th]].



\bibitem{Horndeski:1974wa}
  G.~W.~Horndeski,
  Int.\ J.\ Theor.\ Phys.\  {\bf 10}, 363 (1974).


\bibitem{Deffayet:2011gz}
  C.~Deffayet, X.~Gao, D.~A.~Steer and G.~Zahariade,
  Phys.\ Rev.\ D {\bf 84}, 064039 (2011)
  [arXiv:1103.3260 [hep-th]].


\bibitem{Gleyzes:2014dya}
  J.~Gleyzes, D.~Langlois, F.~Piazza and F.~Vernizzi,
  Phys.\ Rev.\ Lett.\  {\bf 114}, 21, 211101 (2015)
  [arXiv:1404.6495 [hep-th]].




\bibitem{Qiu:2011cy}
  T.~Qiu, J.~Evslin, Y.~F.~Cai, M.~Li and X.~Zhang,
  JCAP {\bf 1110}, 036 (2011)
  [arXiv:1108.0593 [hep-th]].

\bibitem{Qiu:2013eoa}
  T.~Qiu, X.~Gao and E.~N.~Saridakis,
  Phys.\ Rev.\ D {\bf 88}, no. 4, 043525 (2013)
  [arXiv:1303.2372 [astro-ph.CO]].

\bibitem{Cai:2012va}
  Y.~F.~Cai, D.~A.~Easson and R.~Brandenberger,
  JCAP {\bf 1208}, 020 (2012)
  [arXiv:1206.2382 [hep-th]].
%


\bibitem{Wan:2015hya}
  Y.~Wan, T.~Qiu, F.~P.~Huang, Y.~F.~Cai, H.~Li and X.~Zhang,
  JCAP {\bf 1512}, no. 12, 019 (2015)
  [arXiv:1509.08772 [gr-qc]].

\bibitem{Battarra:2014tga}
  L.~Battarra, M.~Koehn, J.~L.~Lehners and B.~A.~Ovrut,
  JCAP {\bf 1407}, 007 (2014)
  [arXiv:1404.5067 [hep-th]].

\bibitem{Koehn:2015vvy}
  M.~Koehn, J.~L.~Lehners and B.~Ovrut,
  Phys.\ Rev.\ D {\bf 93}, no. 10, 103501 (2016)
  [arXiv:1512.03807 [hep-th]].


\bibitem{Libanov:2016kfc}
  M.~Libanov, S.~Mironov and V.~Rubakov,
  arXiv:1605.05992 [hep-th].

\bibitem{Kolevatov:2016ppi}
  R.~Kolevatov and S.~Mironov,
  arXiv:1607.04099 [hep-th].

\bibitem{Kobayashi:2016xpl}
  T.~Kobayashi,
  Phys.\ Rev.\ D {\bf 94}, no. 4, 043511 (2016)
  [arXiv:1606.05831 [hep-th]].


\bibitem{Ijjas:2016vtq}
  A.~Ijjas and P.~J.~Steinhardt,
  arXiv:1609.01253 [gr-qc].


\bibitem{Ijjas:2016tpn}
  A.~Ijjas and P.~J.~Steinhardt,
  arXiv:1606.08880 [gr-qc].




\bibitem{Cheung:2007st}
  C.~Cheung, P.~Creminelli, A.~L.~Fitzpatrick, J.~Kaplan and L.~Senatore,
  JHEP {\bf 0803}, 014 (2008)
  [arXiv:0709.0293 [hep-th]].


\bibitem{Weinberg:2008hq}
  S.~Weinberg,
  Phys.\ Rev.\ D {\bf 77}, 123541 (2008)
  [arXiv:0804.4291 [hep-th]].

\bibitem{Gubitosi:2012hu}
  G.~Gubitosi, F.~Piazza and F.~Vernizzi,
  JCAP {\bf 1302}, 032 (2013)
  [JCAP {\bf 1302}, 032 (2013)]
  [arXiv:1210.0201 [hep-th]].

\bibitem{Gleyzes:2013ooa}
  J.~Gleyzes, D.~Langlois, F.~Piazza and F.~Vernizzi,
  JCAP {\bf 1308}, 025 (2013)
  [arXiv:1304.4840 [hep-th]].



\bibitem{Piazza:2013coa}
  F.~Piazza and F.~Vernizzi,
  Class.\ Quant.\ Grav.\  {\bf 30}, 214007 (2013)
  [arXiv:1307.4350 [hep-th]].

\bibitem{Kase:2014cwa}
  R.~Kase and S.~Tsujikawa,
  Int.\ J.\ Mod.\ Phys.\ D {\bf 23}, no. 13, 1443008 (2015)
  [arXiv:1409.1984 [hep-th]].



\bibitem{Gao:2014soa}
  X.~Gao,
  Phys.\ Rev.\ D {\bf 90}, 081501 (2014)
  [arXiv:1406.0822 [gr-qc]].

\bibitem{Gao:2014fra}
  X.~Gao,
  Phys.\ Rev.\ D {\bf 90}, 104033 (2014)
  [arXiv:1409.6708 [gr-qc]].


\bibitem{Creminelli:2016zwa}
  P.~Creminelli, D.~Pirtskhalava, L.~Santoni and E.~Trincherini,
  arXiv:1610.04207 [hep-th].



%

\bibitem{Cai:2016ldn}
  Y.~Cai, Y.~T.~Wang and Y.~S.~Piao,
  Phys.\ Rev.\ D {\bf 94}, 4, 043002 (2016)
  [arXiv:1602.05431 [astro-ph.CO]].



\bibitem{Easson:2011zy}
  D.~A.~Easson, I.~Sawicki and A.~Vikman,
  JCAP {\bf 1111}, 021 (2011)
  [arXiv:1109.1047 [hep-th]].


\bibitem{Khoury:2001wf}
  J.~Khoury, B.~A.~Ovrut, P.~J.~Steinhardt and N.~Turok,
  Phys.\ Rev.\ D {\bf 64}, 123522 (2001)
  [hep-th/0103239].

\bibitem{Lehners:2007ac}
  J.~L.~Lehners, P.~McFadden, N.~Turok and P.~J.~Steinhardt,
  Phys.\ Rev.\ D {\bf 76}, 103501 (2007)
  [hep-th/0702153 [HEP-TH]].

\bibitem{Buchbinder:2007ad}
  E.~I.~Buchbinder, J.~Khoury and B.~A.~Ovrut,
  Phys.\ Rev.\ D {\bf 76}, 123503 (2007)
  [hep-th/0702154].




\bibitem{Li:2016awk}
  H.~G.~Li, Y.~Cai and Y.~S.~Piao,
  arXiv:1605.09586 [gr-qc].



\end{thebibliography}
\end{document}